\documentclass[aps,showpacs,superscriptaddress,notitlepage]{revtex4-1}
\usepackage{amsmath}
\usepackage{epsfig}
\def\bea#1\eea{\begin{align}#1\end{align}}
\newcommand{\nnu}{\nonumber\\}
\newcommand{\bef}{\begin{figure}[htb]\centering}
\newcommand{\eef}{\end{figure}}

\begin{document}
\title{Multiple scattering effects on heavy meson production  \\ 
 in p+A collisions at backward rapidity  }

\date{\today}

\author{Zhong-Bo Kang}
\email{zkang@lanl.gov}
\affiliation{Theoretical Division, 
                   Los Alamos National Laboratory, 
                   Los Alamos, NM 87545, USA}
                   
\author{Ivan Vitev}
\email{ivitev@lanl.gov}
\affiliation{Theoretical Division, 
                   Los Alamos National Laboratory, 
                   Los Alamos, NM 87545, USA}                   

\author{Enke Wang}
\email{wangek@iopp.ccnu.edu.cn}
\affiliation{Institute of Particle Physics, 
                   Central China Normal University, 
                   Wuhan 430079, China} 

\author{Hongxi Xing}
\email{hxing@lanl.gov}

\affiliation{Theoretical Division, 
                   Los Alamos National Laboratory, 
                   Los Alamos, NM 87545, USA}  
                   
\author{Cheng Zhang}
\email{zhangcheng@iopp.ccnu.edu.cn}
\affiliation{Institute of Particle Physics, 
                   Central China Normal University, 
                   Wuhan 430079, China}

\begin{abstract}
We study the incoherent multiple scattering effects on heavy meson production in the backward rapidity region of p+A collisions within the generalized high-twist factorization formalism. We calculate explicitly the double scattering contributions to the heavy meson differential cross sections by taking into account both initial-state and final-state interactions, and find that these corrections are positive. We further evaluate the nuclear modification factor for  muons that come form the semi-leptonic decays of heavy flavor mesons. Phenomenological applications in d+Au collisions at a center-of-mass energy $\sqrt{s}=200$~GeV at RHIC and in p+Pb collisions at $\sqrt{s}=5.02$~TeV at the LHC are presented. We find that incoherent multiple scattering  can describe rather well the observed nuclear enhancement in the intermediate $p_T$ region for such reactions.
\end{abstract}

\maketitle 

\section{Introduction}
In recent years, the experimental study and theoretical understanding of nuclear effects that affect hadronic observables in proton-nucleus (p+A) collisions has attracted significant attentions~\cite{Adler:2003ii,Adams:2003im,Arsene:2004ux,ALICE:2012mj,Abelev:2014dsa,CMS:2013cka,Perepelitsa:2014xca}. On one hand, quantifying the differences between p+A and p+p collisions can provide a solid baseline for extracting the properties of the quark-gluon plasma (QGP) created in ultra-relativistic nucleus-nucleus (A+A) collisions, where both the initial-state cold nuclear matter effects and final-state hot dense medium effects modify the final-state observables~\cite{Salgado:2011wc}, including heavy flavor~\cite{Sharma:2009hn,Vogt:2010aa,Sharma:2012dy,Arleo:2013zua}. On the other hand, the physics in p+A collisions is interesting in its own right in that it helps illuminate the QCD dynamics of multiple parton interactions~\cite{Vitev:2006bi}, the transport properties of cold nuclear matter~\cite{Accardi:2003jh,Neufeld:2010dz}, the dense gluon structure of the nucleus~\cite{McLerran:1993ni}, and the multi-parton correlations probed by a propagating parton in p+A collisions~\cite{Qiu:2001hj}.

So far, most of the theoretical efforts have been devoted to the study of the nontrivial QCD dynamics in the forward rapidity region, where the parton momentum fraction $x$ in the nucleus is small and the external probe interacts with the partons inside the nucleus {\it coherently}. Furthermore, the parton momentum fraction in the proton is large and the effects of energy loss from  the external probe are amplified. The resulting nuclear suppression of inclusive particle production cross sections relative to the binary collisions scaled p+p baseline~\cite{Adare:2011sc,Braidot:2010ig} has been addressed in the framework of several theoretical formalisms~\cite{Qiu:2004da,Kang:2011bp,Albacete:2010bs,JalilianMarian:2011dt,Kang:2013hta,Kang:2014lha}. In a previous paper we explored a different regime - the backward rapidity region~\cite{Kang:2013ufa}. Specifically, we studied the single inclusive light hadron production in p+A collisions and demonstrated explicitly that in such a regime all  interference Feynman diagrams drop out due to the lack of nuclear-size $A^{1/3}$ enhancement. Thus, only {\it incoherent} multiple scatterings are relevant. We adopted a generalized high-twist factorization formalism to study these incoherent multiple scattering effects. Within such a formalism, multiple parton interactions manifest 
themselves as power-suppressed corrections to the differential cross section and the contributions can be written in terms of high-twist multi-parton correlation functions. For recent progress on the next-to-leading order corrections within this formalism and QCD evolution of the associated high-twist correlation functions, see Refs.~\cite{Kang:2013raa, Xing:2014kpa,Kang:2014ela}. By taking into account both initial-state and final-state interactions, we derived the incoherent double scattering contributions to the differential cross section for single inclusive light hadron production in p+A collisions. We showed  that these  contributions are positive and  lead to nuclear enhancement in the backward rapidity region.

In the current paper we generalize our earlier study to open heavy flavor  production in p+A collisions, and use our results to understand the nuclear enhancement observed in the backward rapidity region  at both RHIC and the LHC. Other approaches in understanding the backward rapidity region include the use of a universal nuclear parton distribution functions (nPDFs), see for example~\cite{Eskola:2009uj}, which appear to give a somewhat unsatisfactory description of the experimental data on heavy flavor decay muons at RHIC~\cite{Adare:2013lkk}. One of the main differences in our approach is that the calculated double scattering contributions are process-dependent, that is {\it non-universal}. We find that, because of the heavy quark mass, the double scattering corrections can no longer be written in the same simple compact form as for  light hadron production. However, as we will show below,  such incoherent double scattering still gives a positive contribution to the heavy meson differential cross section in the backward rapidity region. We calculate the nuclear modification factor for single muons coming from  heavy flavor meson decays, and find that the incoherent double scattering effects can describe rather well the corresponding RHIC and LHC data. We expect that our results will shed light on the origin of cold nuclear matter effects in the backward rapidity region. 

The rest of our paper is organized as follows: in Sec. II we first review the single scattering contribution to single inclusive heavy meson production in p+A collisions. We then extend our previous calculation of the double scattering contribution for light hadron production to heavy meson production. In Sec. III, based on our analytical results, we evaluate the nuclear modification factor for  muons coming form the semi-leptonic heavy flavor decays in d+Au collisions at RHIC and p+Pb collisions at the LHC. We also present comparison to the experimental data. A summary of our paper is given in Sec. IV.

\section{multiple scattering contributions to heavy flavor meson production in p+A collisions}

\subsection{Single scattering contribution}
\label{sec-single}
In this section we consider single inclusive heavy  meson production in p+A collisions,
\bea
p(P') + A(P) \to H(P_h) + X,
\eea
where $H$ represents the observed charm or beauty meson with momentum $P_h$ and mass $m_h$, $P'$ is the momentum of the incoming proton, and $P$ is the momentum per nucleon in the nucleus. In general, the differential cross section for single inclusive particle production in proton-nucleus collisions can be expanded in terms of single scattering, double scattering, and even larger number of scatterings~\cite{Qiu:2001hj}:
\bea
d\sigma_{pA\to HX}=d\sigma_{pA\to HX}^{(S)}+d\sigma_{pA\to HX}^{(D)}+\cdots ,
\eea
where the superscript ``$(S)$'' and ``$(D)$'' represent the contributions of single scattering and double scattering, respectively. The single scattering contribution for heavy meson production at large transverse momentum can be derived within the usual leading-twist perturbative QCD factorization formalism~\cite{Collins:1989gx}, and at leading order in the strong coupling $\alpha_s$ it has the following form:
\bea
E_h\frac{d\sigma^{(S)}}{d^3P_h}=\frac{\alpha_s^2}{s}\sum_{a,b}\int\frac{dz}{z^2}D_{c\to H}(z)\int\frac{dx'}{x'}f_{a/p}(x')
\int\frac{dx}{x}f_{b/A}(x)H^U_{ab\to c}(\hat s,\hat t,\hat u)\delta(\hat s+\hat t+\hat u),
\label{eq-xsing}
\eea
where $\sum_{a,b}$ represents the sum over all parton flavors and the center-of-mass energy squared is $s=(P'+P)^2$. $f_{a/p}(x')$ and $f_{b/A}(x)$ are the usual leading-twist parton distribution functions, and $D_{c\to H}(z)$ is the fragmentation function for a parton $c$ fragmenting into a heavy flavor meson $H$. $H^U_{ab\to c}$ is a short-distance hard-part function for two partons of flavor $a$ and $b$ to produce a parton $c$. It is important to realize that both heavy flavor and light flavor partons (heavy quark, light quark, and gluon) can fragment into the heavy flavor meson. Following Refs.~\cite{Kniehl:2005ej,Kniehl:2012ti}, we take all these contributions into consideration. 

The hard-part functions $H^U_{ab\to c}$ for the light flavor contribution (i.e., parton $c$ is a light quark $q$ or gluon $g$, which then fragments into the heavy meson) are well-known and can be found in Ref.~\cite{Owens:1986mp}. On the other hand, the hard-part functions $H^U_{ab\to c}$ for heavy flavor contribution (i.e., parton $c$ is a heavy quark $Q$) gets contributions from both the light quark-antiquark annihilation $q\bar q\to Q\bar Q$ and gluon-gluon fusion $gg\to Q\bar Q$ subprocesses, as illustrated in Fig.~\ref{fig-single}, and are given by~\cite{Vitev:2006bi,Kang:2008ih}
\bea
H^U_{q\bar q\to Q\bar{Q}}&=\frac{N_c^2-1}{2N_c^2}\frac{\hat t^2+\hat u^2 +2m_c^2\hat s}{\hat s^2},\\
H^U_{gg\to \to Q\bar{Q}}&=\frac{1}{2N_c}\left(\frac{1}{\hat t\hat u}-\frac{2N_c^2}{N_c^2-1}\frac{1}{\hat s^2}\right)
\left(\hat t^2 + \hat u^2 + 4m_c^2\hat s-\frac{4m_c^2\hat s^2}{\hat t\hat u}\right),
\eea
where $N_c=3$ is the number of colors,  and $m_c$ is the mass of the heavy quark $c$ that fragments into the heavy flavor meson $H$. The slightly modified Mandelstam variables $\hat s$, $\hat t$, $\hat u$ are defined at the partonic level as 
\bea
\hat s = (x'P'+xP)^2,
\qquad
\hat t = (x'P'-p_c)^2-m_c^2,
\qquad
\hat u = (xP-p_c)^2-m_c^2,
\label{modMand}
\eea
where $p_c$ is the momentum of the heavy quark $c$, and $\hat s$, $\hat t$, and $\hat u$ satisfies 
$\hat s+\hat t+\hat u=0$ as indicated in Eq. (\ref{eq-xsing}). 
\bef
\psfig{file=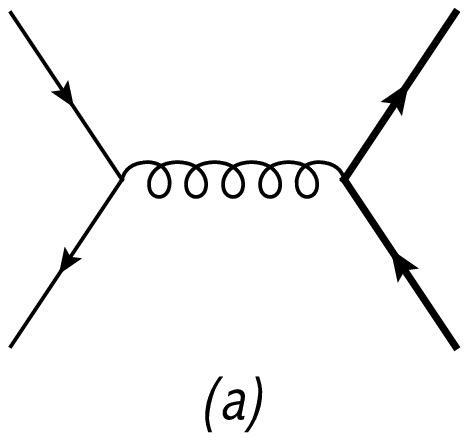, width=1.2in}
\hskip 0.4in
\psfig{file=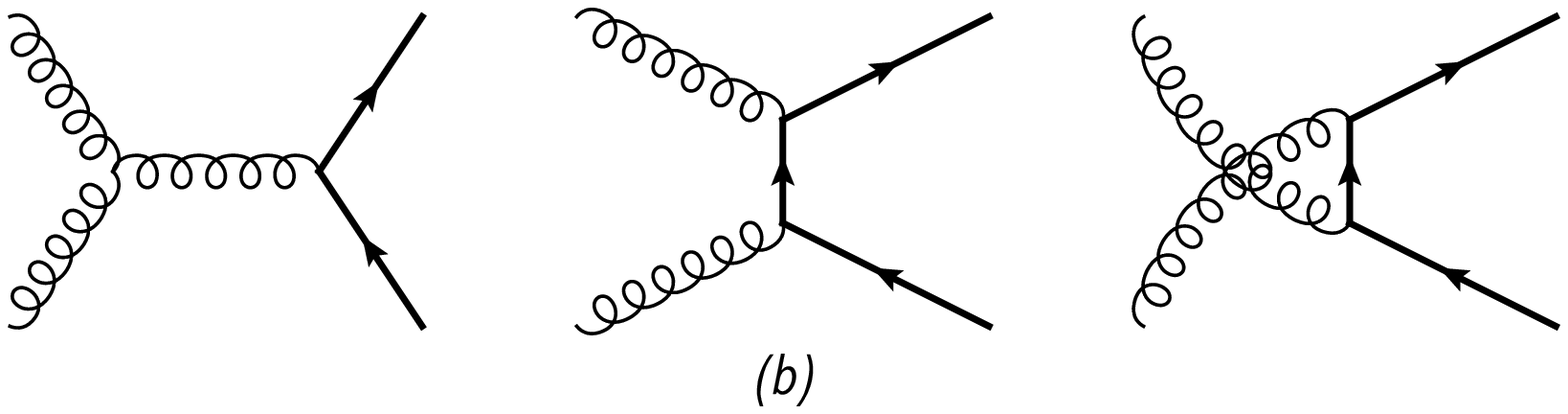, width=4.5in}
\caption{Lowest order Feynman diagrams for heavy quark production at leading twist: (a) light quark-antiquark annihilation $q\bar q\to Q\bar Q$, (b) gluon-gluon fusion $gg\to Q\bar Q$. The thick solid lines represent a heavy quark $Q$.}
\label{fig-single}
\eef

\subsection{Double scattering contributions}
Let us now study the multiple scattering contributions to charm and beauty  meson production. In particular, we focus on the backward rapidity region where the parton momentum fraction in the nucleus is outside the  small-$x$ regime, and {\it incoherent} multiple interactions are important~\cite{Kang:2013ufa}. Within the high-twist collinear factorization formalism~\cite{Luo:1994np}, the first non-trivial multiple scattering (double scattering) contributions are attributed to the twist-4 power-suppressed corrections to the differential cross section. This formalism has been used to study  cold nuclear matter effects in lepton+nucleus and hadron+nucleus collisions, including parton energy loss~\cite{Xing:2011fb,Wang:2001ifa,Guo:2000nz,Zhang:2003wk}, transverse momentum broadening~\cite{Guo:1998rd,Kang:2012am,Kang:2011bp,Xing:2012ii,Kang:2013raa}, and dynamical shadowing~\cite{Qiu:2003vd,Qiu:2004qk,Qiu:2004da,Vitev:2006bi}. The detailed techniques are well explained in our previous paper~\cite{Kang:2013ufa}, where we computed the {\it incoherent} double scattering contributions to the single inclusive {\it light} hadron production in p+A collisions. 

As we have emphasized above, both light and heavy flavor partons can fragment into a heavy meson. For the case when the light flavor parton fragments into a heavy meson, since multiple scattering occurs at the partonic level,  we can immediately obtain the double scattering contributions to the differential cross section by replacing the fragmentation function of light hadron with heavy meson, $D_{c\to h}(z)\to D_{c\to H}(z)$, in the result from our previous paper~\cite{Kang:2013ufa}. For later convenience in the
phenomenological study we write down explicitly the contributions from light flavor fragmentation at twist-4:
\bea
\left.E_h\frac{d\sigma^{(D)}}{d^3P_h}\right|_{\rm light} =&\left(\frac{8\pi^2\alpha_s}{N_c^2-1}\right) 
\frac{\alpha_s^2}{s} \sum_{a,b,c}\int \frac{dz}{z^2} D_{c\to H}(z) \int \frac{dx'}{x'} f_{a/p}(x')
\int \frac{dx}{x} 
\delta(\hat s+\hat t+\hat u)
\nnu
& \times \sum_{i=I, F}
\left[x^2\frac{\partial^2 T^{(i)}_{b/A}(x)}{\partial x^2} 
-x\frac{\partial T^{(i)}_{b/A}(x)}{\partial x} +  T^{(i)}_{b/A}(x)\right] c^{i}
H^{i}_{ab\to cd}(\hat s, \hat t, \hat u),
\label{eq-T4l}
\eea
where the subscript ``light'' denotes the light flavor fragmentation contributions, $\sum_{i=I,F}$ represents the sum over the initial-state and final-state double scattering as explained in \cite{Kang:2013ufa} and below, and the factors $c^{i}$ are given by
\bea
c^{I} =& -\frac{1}{\hat t} - \frac{1}{\hat s},
\label{cI}
\\
c^{F} =&  -\frac{1}{\hat t} - \frac{1}{\hat u},
\label{cF}
\eea
while the hard-scattering functions $H^{i}_{ab\to cd}(\hat s, \hat t, \hat u)$ have the following form:
\bea
H^I_{ab\to c d} =& \left\{
  \begin{array}{l l}
    C_F H^U_{ab\to c d} & \quad \text{a\ =\ quark}\\
     \\
    C_A H^U_{ab\to c d} & \quad \text{a\ =\ gluon}\\
  \end{array} \right.  \; , 
  \label{HI}
  \\
  \nnu
H^F_{ab\to c d} =& \left\{
  \begin{array}{l l}
    C_F H^U_{ab\to c d} & \quad \text{c\ =\ quark}\\
     \\
    C_A H^U_{ab\to c d} & \quad \text{c\ =\ gluon}\\
  \end{array} \right.   \; .
  \label{HF}
\eea
The relevant initial-state correlation functions are the so-called quark-gluon correlation function $T_{q/A}^{(I)}(x)$ and  gluon-gluon correlation function $T_{g/A}^{(I)}(x)$, and they have the following definitions \cite{Kang:2013ufa}:
\bea
T_{q/A}^{(I)}(x) = &
 \int \frac{dy^{-}}{2\pi}\, e^{ixP^{+}y^{-}}
 \int \frac{dy_1^{-}dy_{2}^{-}}{2\pi} \,
      \theta(y^{-}-y_{1}^{-})\,\theta(-y_{2}^{-}) 
     \frac{1}{2}
     \langle P |F_{\alpha}^{\ +}(y_{2}^{-})\bar{\psi}_{q}(0)
                  \gamma^{+}\psi_{q}(y^{-})F^{+\alpha}(y_{1}^{-}) |P \rangle\, ,
\label{TqA}
\\
T_{g/A}^{(I)}(x) = &
 \int \frac{dy^{-}}{2\pi}\, e^{ix P^{+}y^{-}}
 \int \frac{dy_1^{-}dy_{2}^{-}}{2\pi} \,
      \theta(y^{-}-y_{1}^{-})\,\theta(-y_{2}^{-}) 
\frac{1}{x P^+}\,
\langle P| F_\alpha^{~+}(y_2^-)
F^{\sigma+}(0)F^+_{~\sigma}(y^-)F^{+\alpha}(y_1^-)|P\rangle\, .
\label{TgA}
\eea
On the other hand, the relevant final-state correlation functions $T^{(F)}_{q,g/A}(x)$ are the same as 
$T^{(I)}_{q,g/A}(x)$, except for the $\theta$-functions that are replaced as follows~\cite{Kang:2011bp,Kang:2008us,Xing:2012ii}
\bea
\theta(y^{-}-y_{1}^{-})\,\theta(-y_{2}^{-}) 
\to
\theta(y_{1}^{-}-y^{-})\,\theta(y_{2}^{-}).
\eea
One interesting feature of the light flavor contributions in Eq.~\eqref{eq-T4l} is that the second derivative, the first derivative, and the non-derivative  $T_{q,g/A}^{I,F}(x)$  terms share the common hard-part function, and  have a very compact simple form. We will see below that such a feature will no longer hold for the heavy flavor contributions because of the mass terms. 

We now turn to the double scattering contribution for  the prompt heavy quark final states, which is the main new result of our analytical calculations. In other words, we study both the initial-state and final-state double scattering corrections to the partonic processes $q\bar q\to Q\bar Q$ and $gg\to Q\bar Q$. The relevant Feynman diagrams are shown in Figs.~\ref{fig-q+gq} and \ref{fig-g+gg}, respectively. 
Here, the initial-state multiple scattering represents the situations where the incoming parton from the proton undergoes multiple interactions with the soft partons inside the nucleus {\it before} the hard collisions, while final-state multiple scattering stands for the situations where the leading outgoing parton undergoes multiple interactions in the large nucleus {\it after} the hard collisions. 

\bef
\psfig{file=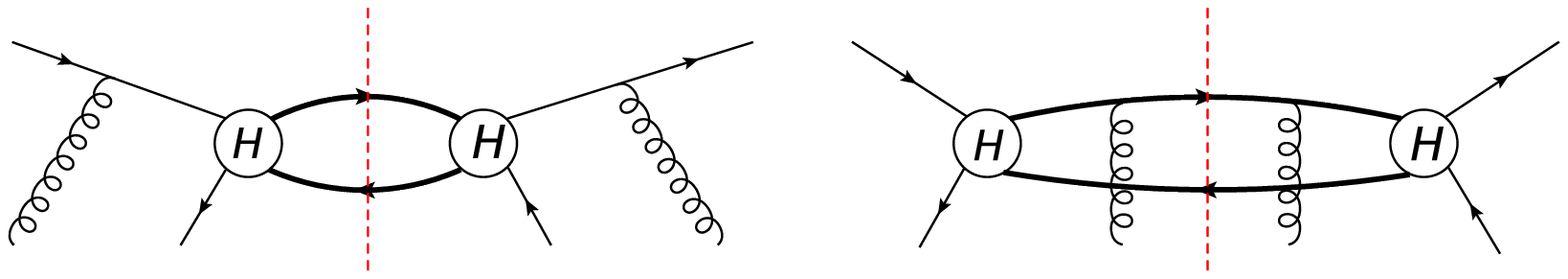, width=6in}
\caption{The central-cut diagrams for initial-state (left) and final-state (right) double scatterings in quark-antiquark annihilation process. The ``H''-blobs represent the hard $q\bar q\to Q\bar Q$ processes as shown in Fig.~\ref {fig-single}(a).}
\label{fig-q+gq}
\eef

Since all techniques for computing the double scattering corrections  are the same as those for light hadron production, we skip all the details of the calculations and make only a few remarks about our derivation. First, besides those Feynman diagrams presented in Figs.~\ref{fig-q+gq} and \ref{fig-g+gg} in which there is only one additional rescattered gluon in each side of the unitarity cut line in a classical double scattering picture, we also include the diagrams  in which two rescattered gluons are on the same side of the unitarity cut line, representing the interferences between single and triple scatterings. These interference diagrams give the so-called ``contact'' contributions (i.e., have no nuclear-size $A^{1/3}$ enhancement) when combined with the central-cut diagrams in the large-$x$ regime and  will be neglected.  For more details, see \cite{Kang:2013ufa}. Because of this, our final results only depend on the four-parton correlation functions associated with the central-cut diagrams and, thus, represent the classical incoherent scattering regime. Second, the Feynman diagrams in which the rescattering happens between the unobserved outgoing parton $d$ and the nuclear medium also leads to ``contact'' contributions and is also neglected.  We finally obtain the following result for the double scattering contributions to the $ q\bar q\to Q\bar Q$ process,
\bea
\left. E_h\frac{d\sigma^{(D)}}{d^3P_h}\right|_{q\bar{q}\to Q\bar{Q}}=&\frac{8\pi^2\alpha_s}{N_c^2-1}\frac{\alpha_s^2}{s}\sum_{q}\int\frac{dz}{z^2}D_{Q\to H}(z)\frac{dx'}{x'}f_{q/p}(x')
\int\frac{dx}{x}H^U_{q\bar{q}\to Q\bar{Q}}(\hat s,\hat t,\hat u)\delta(\hat s+\hat t+\hat u)
\nnu
&\times\sum_{i=I,F}\left(x^2\frac{\partial^2T_{\bar{q}/A}(x)}{\partial x^2}c_2^{qi}
-x\frac{\partial T_{\bar{q}/A}(x)}{\partial x}c_1^{qi}+T_{\bar{q}/A}(x)c_0^{qi}\right),
\label{eq-T4q}
\eea
where the double scattering hard-part functions are proportional to the single-scattering one $H^U_{q\bar q\to Q\bar Q}$ with the pre-factor coefficients $c_{0,1,2}^{qi}$ given by
\bea
c_2^{qI}&=C_F\left[-\frac{1}{\hat t}-\frac{1}{\hat s}-\frac{m_c^2}{\hat t^2}\right],
\\
c_1^{qI}&=C_F\left[-\frac{1}{\hat t}-\frac{1}{\hat s}-2\frac{m_c^2}{\hat t^2}\frac{(\hat t-\hat u)^2+4m_c^2\hat s}
{2m_c^2\hat s+\hat t^2+\hat u^2}\right],
\\
c_0^{qI}&=C_F\left[-\frac{1}{\hat t}-\frac{1}{\hat s}-2\frac{m_c^2}{\hat t^2}\frac{(\hat t-\hat u)^2-\hat t\hat u+6m_c^2\hat s}
{2m_c^2\hat s+\hat t^2+\hat u^2}\right],
\\
c_2^{qF}&=C_F\left[-\frac{1}{\hat t}-\frac{1}{\hat u}-\frac{m_c^2\hat s^2}{\hat t^2\hat u^2}\right],
\\
c_1^{qF}&=C_F\left[-\frac{1}{\hat t}-\frac{1}{\hat u}-2\frac{m_c^2\hat s^2}{\hat t^2\hat u^2}\frac{(\hat t-\hat u)^2+4m_c^2\hat s}
{2m_c^2\hat s+\hat t^2+\hat u^2}\right],
\\
c_0^{qF}&=C_F\left[-\frac{1}{\hat t}-\frac{1}{\hat u}-2\frac{m_c^2\hat s^2}{\hat t^2\hat u^2}\frac{(\hat t-\hat u)^2-\hat t\hat u+6m_c^2\hat s}
{2m_c^2\hat s+\hat t^2+\hat u^2}\right],
\eea
where we have the color factor $C_F$ for both initial-state and final-state double scattering, representing the color interaction strength between the incoming light quark $q$ (or outgoing heavy quark $Q$) and the soft partons in the nucleus. We also recall that for prompt heavy quark final states the definition of the Mandelstam variables is given in Eq.~(\ref{modMand}).
It is instructive to notice that because the mass  generates extra terms in the above pre-factor coefficients, we do not have the simple compact form as in the light flavor fragmentation. As a consistency check, if one takes heavy quark mass $m_c\to 0$ limit, we recover the same result for $q\bar q\to q'\bar q'$ \cite{Kang:2013ufa}. 

\bef
\psfig{file=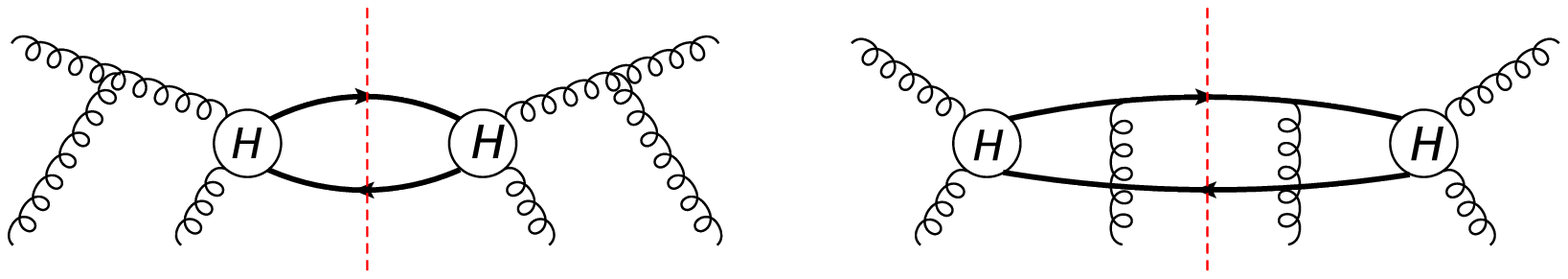, width=6in}
\caption{The central-cut diagrams for initial-state (left) and final-state (right) double scattering in the gluon-gluon fusion process. The ``H''-blobs represent the hard $gg\to Q\bar Q$ processes as shown in Fig. \ref {fig-single}(b).}
\label{fig-g+gg}
\eef
On the other hand, the double scattering contributions to the $gg\to Q\bar Q$ fusion process are given in Fig. \ref{fig-g+gg}. The calculations are similar, and the final result reads
\bea
\left. E_h\frac{d\sigma^{(D)}}{d^3P_h}\right|_{gg\to Q\bar{Q}}=&\frac{8\pi^2\alpha_s}{N_c^2-1}\frac{\alpha_s^2}{s}\int\frac{dz}{z^2}D_{Q\to H}(z)\frac{dx'}{x'}f_{g/p}(x')
\int\frac{dx}{x}H^U_{gg\to Q\bar{Q}}(\hat s,\hat t,\hat u)\delta(\hat s+\hat t+\hat u)\nnu
&\times\sum_{i=I,F}\left(x^2\frac{\partial^2T_{g/A}(x)}{\partial x^2}c_2^{gi}
-x\frac{\partial T_{g/A}(x)}{\partial x}c_1^{gi}+T_{g/A}(x)c_0^{gi}\right),
\label{eq-T4g}
\eea
where again the double scattering hard-part functions are proportional to the single scattering one $H^U_{gg\to Q\bar{Q}}$ with the following pre-factor coefficients:
\bea
c_2^{gI}&=C_A\left[-\frac{1}{\hat t}-\frac{1}{\hat s}-\frac{m_c^2}{\hat t^2}\right],
\\
c_1^{gI}&=C_A\left[-\frac{1}{\hat t}-\frac{1}{\hat s}+2\frac{m_c^2}{\hat t^2}
\frac{12m_c^4\hat s^3-16m_c^2\hat s^2\hat t\hat u+\hat t\hat u\left(\hat t^3+3\hat s\hat t\hat u+\hat u^3\right)}
{\hat s\left(-4m_c^4\hat s^2+4m_c^2\hat s\hat t\hat u+\hat t^3\hat u+\hat t\hat u^3\right)}\right],
\\
c_0^{gI}&=C_A\left[-\frac{1}{\hat t}-\frac{1}{\hat s}+2\frac{m_c^2}{\hat t^2}
\frac{24m_c^4\hat s^3-28m_c^2\hat s^2\hat t\hat u-\hat s\hat t\hat u\left(\hat t^2-6\hat t\hat u+\hat u^2\right)}
{\hat s\left(-4m_c^4\hat s^2+4m_c^2\hat s\hat t\hat u+\hat t^3\hat u+\hat t\hat u^3\right)}\right],
\\
c_2^{gF}&=C_F\left[-\frac{1}{\hat t}-\frac{1}{\hat u}-\frac{m_c^2\hat s^2}{\hat t^2\hat u^2}\right],
\\
c_1^{gF}&=C_F\left[-\frac{1}{\hat t}-\frac{1}{\hat u}+2\frac{m_c^2\hat s}{\hat t^2\hat u^2}
\frac{12m_c^4\hat s^3-16m_c^2\hat s^2\hat t\hat u+\hat t\hat u\left(\hat t^3+3\hat s\hat t\hat u+\hat u^3\right)}
{-4m_c^4\hat s^2+4m_c^2\hat s\hat t\hat u+\hat t^3\hat u+\hat t\hat u^3}\right],
\\
c_0^{gF}&=C_F\left[-\frac{1}{\hat t}-\frac{1}{\hat u}+2\frac{m_c^2\hat s}{\hat t^2\hat u^2}
\frac{24m_c^4\hat s^3-28m_c^2\hat s^2\hat t\hat u-\hat s\hat t\hat u\left(\hat t^2-6\hat t\hat u+\hat u^2\right)}
{-4m_c^4\hat s^2+4m_c^2\hat s\hat t\hat u+\hat t^3\hat u+\hat t\hat u^3}\right].
\eea
Here, we have the color factor $C_A$ ($C_F$) for initial-state (final-state) double scattering, representing the color interaction strength between the incoming gluon $g$ (outgoing heavy quark $Q$) and the soft partons in the nucleus. Again, we can not express the final results  in the most compact possible form due to the finite mass correction terms. If we take $m_c\to 0$, however, we recover the same result for $gg\to q\bar q$ \cite{Kang:2013ufa}. 

Combining the  cross sections in Eq.~\eqref{eq-T4l}, \eqref{eq-T4q},  and \eqref{eq-T4g} with the corresponding hard-part functions, we have the final result for the double scattering contribution to heavy meson production in p+A collisions as
\bea
E_h\frac{d\sigma^{(D)}}{d^3P_h}=\left. E_h\frac{d\sigma^{(D)}}{d^3P_h}\right|_{\rm light}+\left. E_h\frac{d\sigma^{(D)}}{d^3P_h}\right|_{q\bar q\to Q\bar{Q}}+\left. E_h\frac{d\sigma^{(D)}}{d^3P_h}\right|_{gg\to Q\bar{Q}},
\label{eq-xdouble}
\eea
which will be used in our phenomenological studies in the next section. It is important to emphasize again that only the central-cut Feynman diagrams contribute in the backward region (outside small-$x$). In other words, the {\it incoherent} double scattering contributions in Eq.~\eqref{eq-xdouble} has no interference effects and thus are expected to give positive contributions to the heavy meson cross sections in the backward rapidity region in p+A  collisions, as we will show in the next section.

\section{Phenomenology}
In this section we  present phenomenological applications of our analytic results. We  study  heavy meson production in p+A collisions at both RHIC and LHC energies. 

\bef
\psfig{file=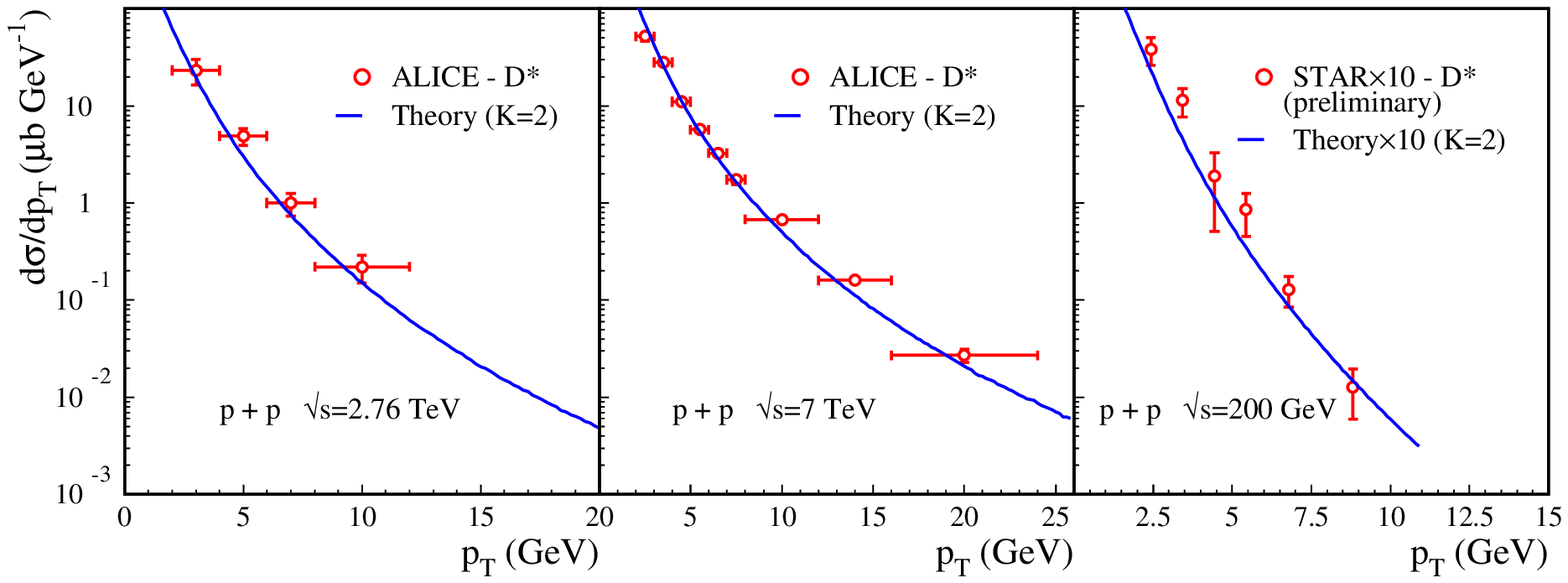, width=0.9\textwidth}
\caption{Comparison of cross section simulations to experimental data for $D^{*+}$ meson production in p+p collisions. Left: LHC case at $\sqrt{s}=2.76$ TeV with rapidity $|y|<0.5$ \cite{Abelev:2012vra}; Middle: LHC case with $\sqrt{s}=7$ TeV with rapidity $|y|<0.5$ \cite{ALICE:2011aa}; Right: RICH case at $\sqrt{s}=200$ GeV with rapidity $|y|<1$ \cite{Adamczyk:2012af,ye:2014}. We set $\mu = m_T$ and  use a same $K$-factor, $K=2$ for both RHIC and LHC energies.}
\label{ALICE_pp_2760}
\eef
We first show  numerical results for heavy meson production in p+p collisions, in which only single scattering contributions are relevant and we start from the  cross section in Eq.~\eqref{eq-xsing}. We use CTEQ6L1 parton distribution functions \cite{Pumplin:2002vw} and the KKKS parametrization for heavy meson fragmentation functions \cite{Kneesch:2007ey}. We choose the factorization and renormalization scales to be equal throughout the numerical study  and set $\mu \sim m_T=\sqrt{m_c^2+p_{h\perp}^2}$ with $m_c = 1.5$ GeV for $D$ meson production. In order to account for  higher order QCD contributions, we include a phenomenological $K$-factor. We find that with a same 
value of the $K$-factor, $K=2$, our leading order formalism describes reasonably well the data from both RHIC at center-of-mass energy $\sqrt{s}=200$~GeV and LHC at $\sqrt{s}=2.76,~7$~TeV for $D$-meson productions. In the left and middle panels of Fig.~\ref{ALICE_pp_2760}  we compare our calculations to LHC data for $D^{*+}$ meson production at $\sqrt{s}=2.76$~TeV \cite{Abelev:2012vra} and $\sqrt{s}=7$~TeV \cite{ALICE:2011aa}, respectively. We also consider RHIC $D^{*+}$ meson production at $\sqrt{s}=200$ GeV, shown in the right panel. Similar agreements are also found for $D^0$ and $D^+$ data. This provides us with a reasonable p+p baseline for our study of heavy meson production in p+A collisions. To this order, any remaining discrepancy in the overall normalization of the cross section will cancel out in the nuclear modification ratio discussed below.     

Keeping in mind that  the Cronin-like enhancement can be considerable for both open heavy flavor and 
quarkonia~\cite{Sharma:2009hn,Sharma:2012dy}, we turn our attention to the study of nuclear effects on heavy meson production in the backward  rapidity region in p+A collisions. As we have emphasized in the last section, this is the region where the {\it incoherent} multiple scattering are relevant and nuclear enhancement (because of the positive contributions from incoherent double scatterings) should be expected. Our numerical simulations below confirm that this is indeed the case. The nuclear effect is usually quantified by the nuclear modification factor $R_{pA}$ defined as follows:
\bea
R_{pA} = \frac{1}{\langle N_{\rm coll}\rangle}
\left.
E_h\frac{d\sigma_{pA}}{d^3P_h} \right/E_h\frac{d\sigma_{pp}}{d^3P_h},
\label{eq-def}
\eea
where $\langle N_{\rm coll}\rangle$ is the average number of binary collisions. $R_{pA}$ is defined such that the deviation from unity reveals the presence of non-trivial nuclear effects in p+A collisions. In our formalism, the denominator in Eq.~\eqref{eq-def} represents the heavy meson production cross section in p+p collisions, as given by Eq.~\eqref{eq-xsing}. On the other hand, the numerator represents the heavy meson cross section in p+A collisions, which receives the contributions from both single and double scatterings, i.e., given by the sum of Eq.~\eqref{eq-xsing} and \eqref{eq-xdouble}.  To numerically evaluate the double scattering contributions, the only unknown ingredients in our formalism are the twist-4 quark-gluon and gluon-gluon correlation functions. These functions represent non-perturbative properties of the nuclear medium, and should in principle be extracted from the experimental data. In Refs.~\cite{Kang:2011bp, Qiu:2003vd}, they have been parametrized as 
\bea
\frac{4\pi^2\alpha_s}{N_c}\,T_{q,g/A}^{(I)}(x)=\frac{4\pi^2\alpha_s}{N_c}\,T_{q,g/A}^{(F)}(x)
= \xi^2 \left(A^{1/3}-1\right) f_{q,g/A}(x),
\label{eq:ht}
\eea
where $\xi^2$ is a universal quantity, representing a characteristic scale and the strength of parton multiple scattering. At tree level, $\xi^2$ can be treated as a fixed number. High order corrections may lead to residual energy and factorization scale dependence of $\xi^2$ after the leading dependence captured by  $f_{q,g/A}(x)$  is taken into account,  see~\cite{Kang:2013raa}.  $\xi^2=0.09-0.12$ GeV$^2$ was extracted from deep inelastic scattering data in e+A collisions~\cite{Qiu:2003vd}, which has been used to describe successfully the nuclear suppression of single inclusive hadron production~\cite{Qiu:2004da} and the di-hadron transverse momentum imbalance and correlations in d+Au collisions at forward rapidities at RHIC $\sqrt{s}=200$ GeV~\cite{Kang:2011bp}. For the purpose of numerical study below, we will use the same value $\xi^2=0.09-0.12$ GeV$^2$. 

\bef
\psfig{file=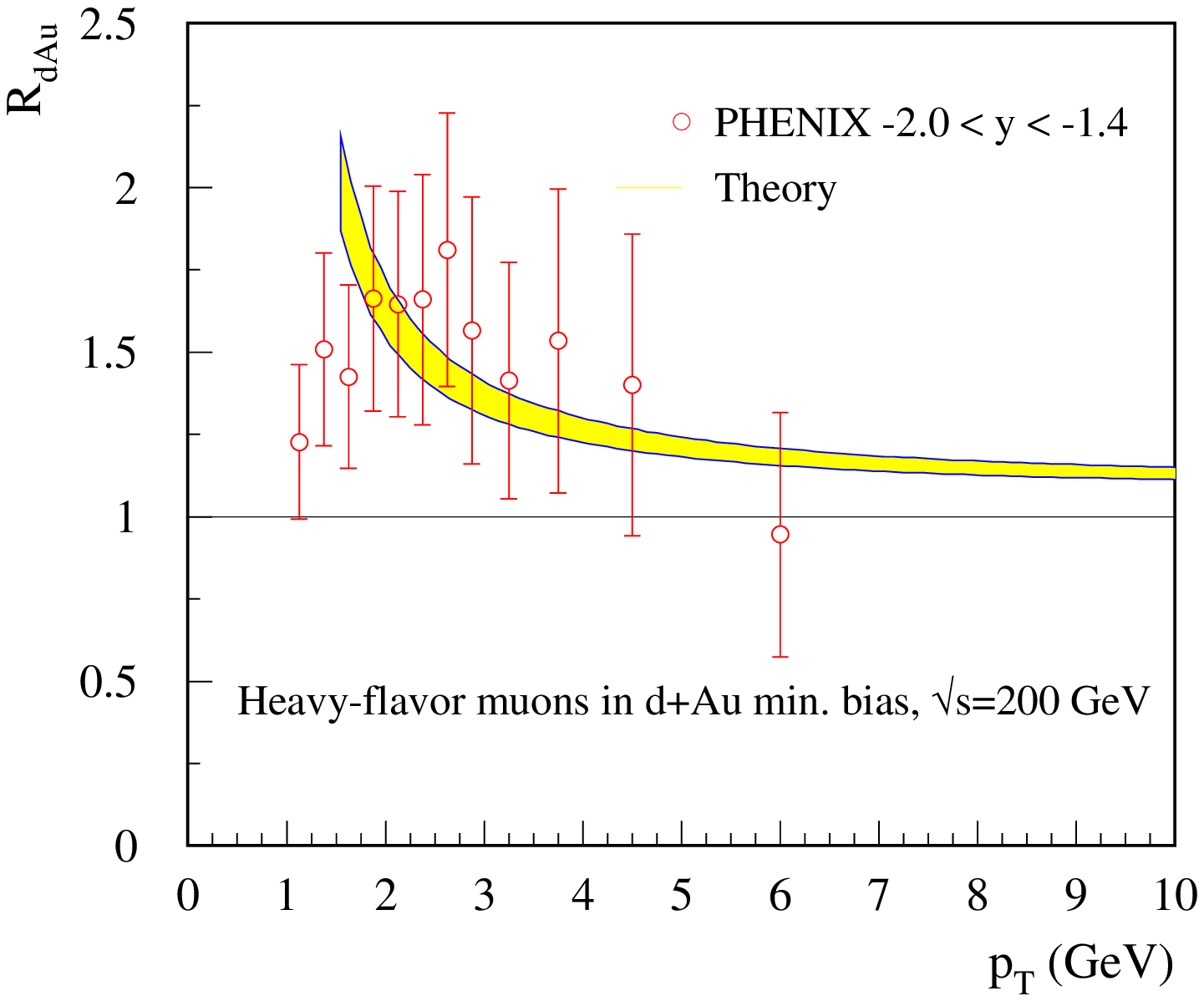, width=0.45\textwidth}
\hskip 0.2in
\psfig{file=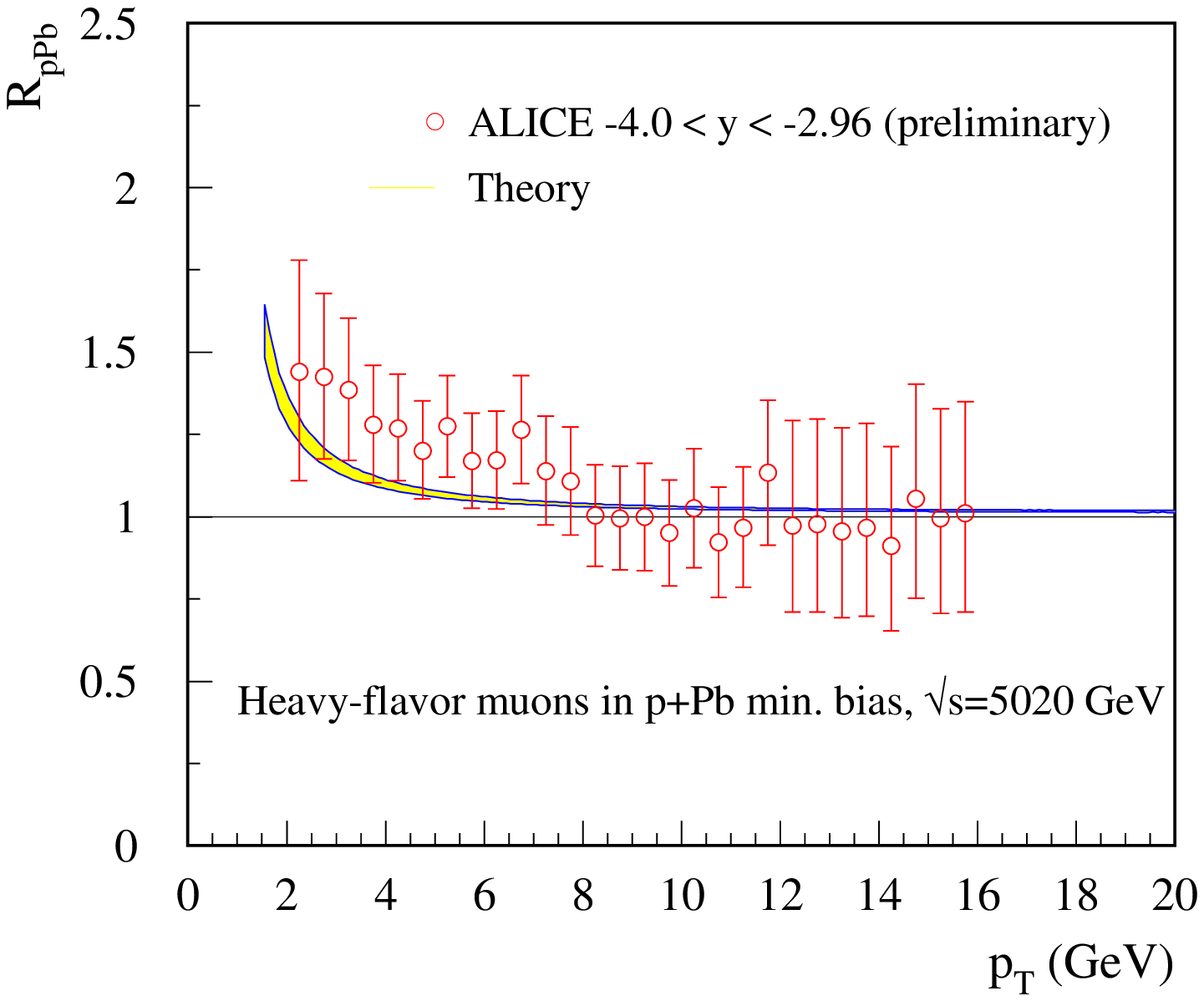, width=0.45\textwidth}
\caption{Comparison of our results to experimental data on the  nuclear modification factor of  muons coming form open heavy flavor decay. Left: RHIC at $\sqrt{s}=200$ GeV with $-2<y<-1.4$ in d+Au collisions \cite{Adare:2013lkk}; Right: LHC at $\sqrt{s}=5.02$ TeV with $-4<y<-2.96$ in p+Pb collisions \cite{Li:2014dha}. We set $\mu = m_T$, and the band corresponds to $\xi^2 = 0.09 - 0.12$ GeV$^2$.}
\label{fig-R_pA}
\eef
In Fig.~\ref{fig-R_pA} we plot the nuclear modification factors $R_{pA}$ for  muons coming form the semi-leptonic open heavy flavor decays in the backward rapidity region in minimum bias d+Au (or p+Pb) collisions   as a function of the muon transverse momentum $p_{T}$. We compare to experimental data from both RHIC~\cite{Adare:2013lkk} (left) and LHC (right)~\cite{Li:2014dha} energies. As in 
Ref.~\cite{Sharma:2009hn}, we use the PYTHIA event generator \cite{Sjostrand:2006za} to simulate the full kinematics of Dalitz decays of heavy mesons in both p+p and p+A collisions. RHIC data is from the PHENIX collaboration at $\sqrt{s}=200$ GeV and rapidity $-2<y<-1.4$ \cite{Adare:2013lkk}. LHC data is from the ALICE collaboration at $\sqrt{s}=5.02$ TeV and rapidity $-4<y<-2.96$ \cite{Li:2014dha}. 
As anticipated,  the incoherent double scatterings give positive contributions to the heavy meson cross section, which lead to a Cronin-like enhancement in the intermediate $p_T$ region. Such an enhancement disappears at large $p_T$, simply because of the high-twist nature of the double scatterings in this framework, i.e. the twist-4 contributions is power suppressed  $\sim 1/p_T^2$. On the other hand, it increases at the low $p_T$ region (even slightly above the RHIC data), because of the same reason, as well as due to the theoretical uncertainty of the $p+p$ baseline in the choice of factorization scale. 
Our numerical calculations give a reasonable description of both RHIC and LHC data, though slightly below the LHC data. Such a slight difference could be further studied in the future. On the experimental side there is no baseline measurements for heavy meson production in p+p collisions at the same center-of-mass energy, $\sqrt{s}=5.02$ TeV, and this can introduce uncertainties in the determined nuclear modification factor $R_{pA}$. On the theory side, one could study the energy/scale dependence of the parameter $\xi^2$, as well as other nuclear matter effects.

\bef
\psfig{file=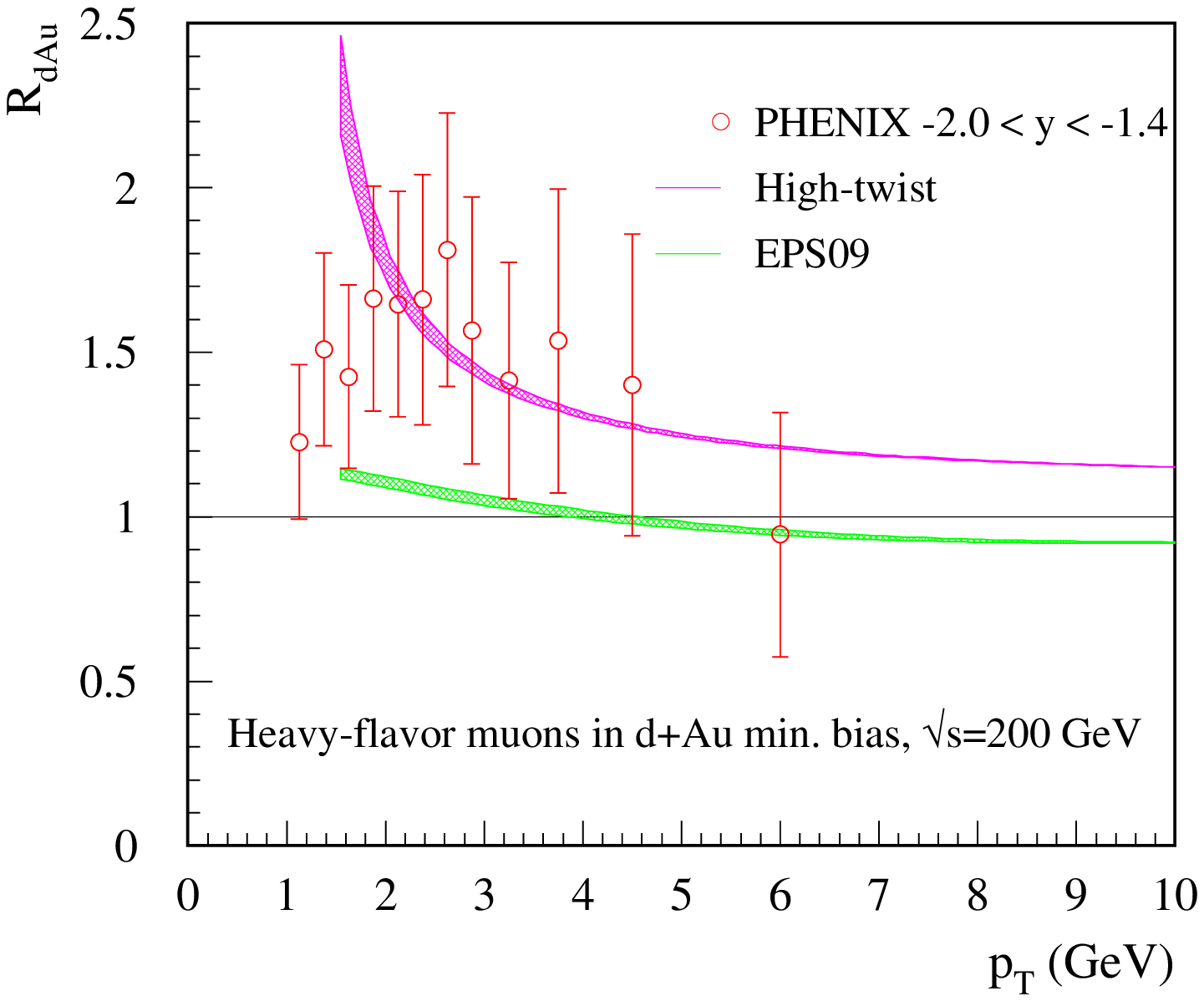, width=0.45\textwidth}
\hskip 0.2in
\psfig{file=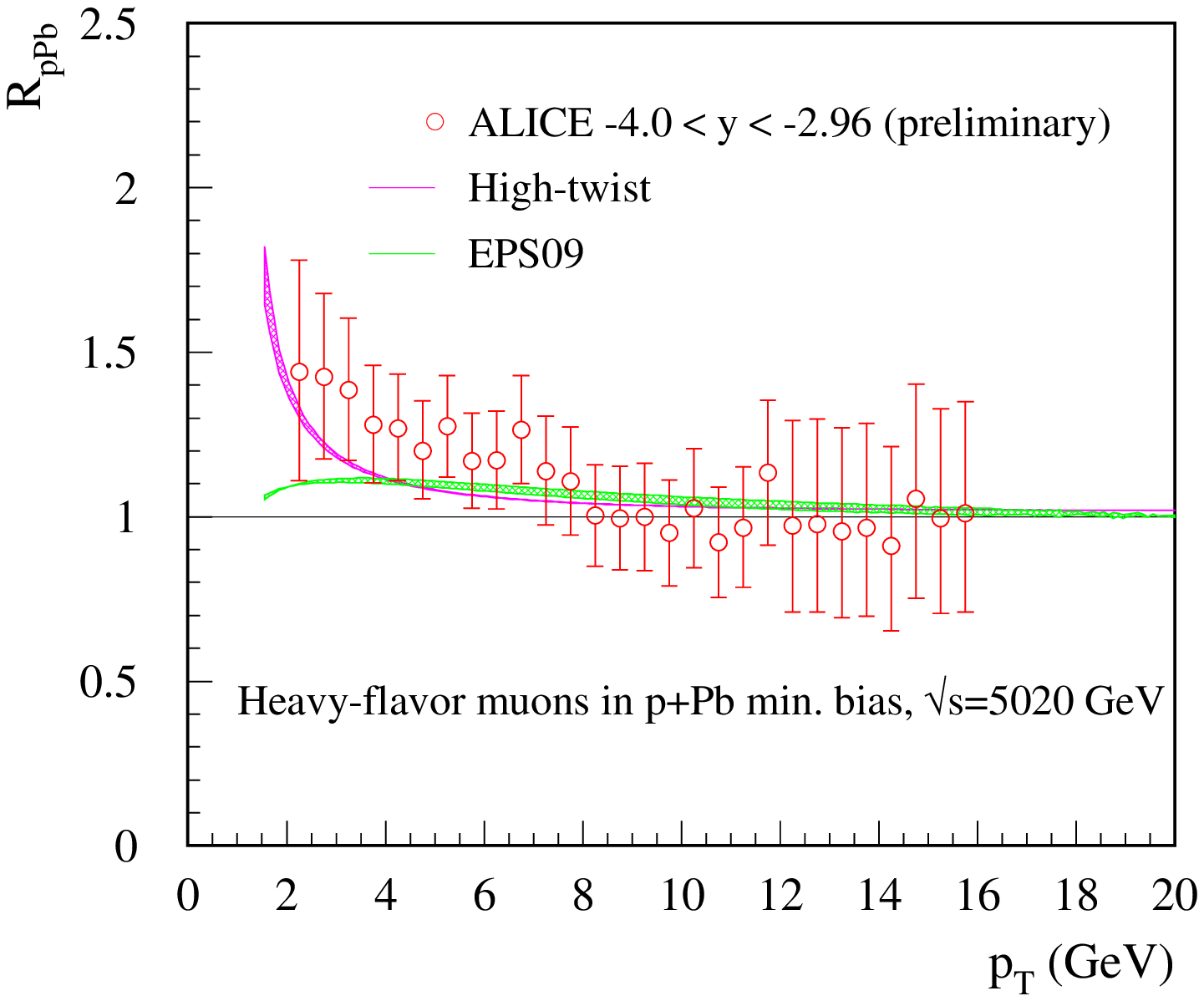, width=0.45\textwidth}
\caption{Comparison of our results (purple band) to the calculation by using nPDFs-EPS09 (green band) on the  nuclear modification factor of  muons coming form open heavy flavor decay at both RHIC (left) and LHC (right). The uncertainty bands correspond to the scale choice $m_T < \mu < 2m_T$. In the high-twist calculation, we set $\xi^2$ = 0.12 GeV$^2$.}
\label{fig-R_pA_nPDFs}
\eef

As pointed out in the introduction, other approaches, such as the use of universal nPDFs, are also considered in the literature. In Fig.~\ref{fig-R_pA_nPDFs} we present a comparison between our approach and the nPDFs approach (using EPS09 parametrization~\cite{Eskola:2009uj}). As can be seen from Fig. \ref{fig-R_pA_nPDFs} (left), the result from EPS09 cannot describe the experimental data on heavy flavor decay muons at RHIC energy~\cite{Adare:2013lkk}. For LHC energy, the results from high-twist and EPS09 are consistent at large $p_T$ region, and disagree in the low $p_T$ region. As we have clarified already, the main difference between our approach and the one that uses nPDFs is that the nuclear modification in high-twist formalism comes from the double scattering contributions, which are process-dependent, that is {\it non-universal}. However, the latter comes from the {\it universal} nPDFs. Therefore, the agreement between our results and experimental data could indicate the non-universality (process dependence) of cold nuclear matter effects. 

We would also like to point out that our calculations are restricted to LO, while higher order contributions are accounted for by implementing a phenomenological $K$-factor. This introduces uncertainty in the theoretical calculation of the differential cross sections. As one can see from the uncertainty band in Fig. \ref{fig-R_pA_nPDFs}, the effect of scale variation $m_T < \mu < 2m_T$ in the nuclear modification ratio is greatly reduced and smaller than the uncertainty due to the choice of $\xi^2$. 

\section{Summary and discussion}
In this paper, we studied the effect of multiple scattering on heavy flavor meson production in p+A collisions. We concentrated on the backward rapidity regime, where the parton momentum fraction $x$ in the nucleus is relatively large and the multiple scattering between the probe and the nuclear medium is {\it incoherent}. Within the high-twist factorization formalism, we evaluated the initial-state and final-state interactions relevant to heavy meson production in p+A collisions. We found that  our final results depend on both the twist-4 quark-gluon and gluon-gluon correlation functions.  Using the existing parametrization for these correlation functions, we calculated numerically the double scattering contribution to the differential cross section of heavy mesons in the kinematic region relevant to both RHIC and LHC experiments. Our simulations  describe quite  well the nuclear modification factor for  muons coming from the semi-leptonic decays of heavy flavor mesons in the backward rapidity region in p+A collisions, where a Cronin-like enhancement is observed experimentally. This feature is understood  as the {\it incoherent} multiple scattering of hard partons in the large nucleus. We conclude that the backward rapidity measurement provide a unique opportunity to investigate the perturbative QCD dynamics in a region that has so far not received adequate attention and to help further  constrain the properties of cold nuclear matter. 

It is important to emphasize that the incoherent multiple scatterings for heavy meson production in p+A collisions involve both initial-state and final-state interactions. To disentangle these two multiple scattering effects, it is instructive to consider processes involving a photon either in the initial state or final state of the process, as  shown in Ref. \cite{Kang:2013ufa}. With the advent of a future electron-ion collider, it will be interesting to study heavy meson production in e+A collisions. This process can provide us with a clean channel to investigate purely the final-state multiple scattering effect. Such a study could also help  clarify the non-universality of nuclear effects in different processes due to the process dependent hard part coefficient \cite{Xing:2012ii}, and test the predictive power of the high twist formalism due to the universality of the twist-4 parton-parton correlation functions. We leave this study for future work.

\section*{Acknowledgments}
We thank X.~Dong and M.~Mustafa for providing the STAR collaboration $D$-meson preliminary data in p+p collisions, and S.~Li for helpful discussions on the ALICE data. This work is supported by the National Natural Science Foundation of China under Grant Nos. 11435004?11221504, the Major State Basic Research Development Program in China (No. 2014CB845404), by US Department of Energy under contract Nos. 2012LANL7033, 2012LALN4005, Office of Science, Office of Nuclear Physics and in part by the LDRD program at LANL under contract No. 2013783PRD2.

\end{document}